\documentclass[twocolumn,superscriptaddress,amsmath,amssymb,aps,prx]{revtex4-1}

\usepackage{graphicx}
\usepackage[space]{grffile}
\usepackage{dcolumn}
\usepackage{bm}
\usepackage{subfigure}
\usepackage{mathrsfs}
\usepackage{booktabs}
\usepackage{notoccite}
\usepackage{float}
\usepackage{placeins}
\usepackage{siunitx}

\begin{document}
	
	\title{A new symmetry-based framework for discovering minimum energy pathways}
	
	\author{Jason M. Munro}
	\email{munrojm@psu.edu}
	\affiliation{Department of Materials Science and Engineering, Materials Research Institute, and Penn State Institutes of Energy and the Environment, The Pennsylvania State University, University Park, PA 16802, USA}
	
	\author{Hirofumi Akamatsu}
	\affiliation{Department of Applied Chemistry, Kyushu University, Fukuoka 819-0395, Japan}
	
	\author{Haricharan Padmanabhan}
    \author{Vincent S. Liu}
    \author{Yin Shi}
    \author{Long-Qing Chen}
	\affiliation{Department of Materials Science and Engineering, Materials Research Institute, and Penn State Institutes of Energy and the Environment, The Pennsylvania State University, University Park, PA 16802, USA}
	
	\author{Brian K. VanLeeuwen}
	\affiliation{Two Sigma Investments, New York, New York 10013, USA}
	
	\author{Ismaila Dabo}
	\affiliation{Department of Materials Science and Engineering, Materials Research Institute, and Penn State Institutes of Energy and the Environment, The Pennsylvania State University, University Park, PA 16802, USA}
	
	\author{Venkatraman Gopalan}
	\affiliation{Department of Materials Science and Engineering, Materials Research Institute, and Penn State Institutes of Energy and the Environment, The Pennsylvania State University, University Park, PA 16802, USA}
	\affiliation{Department of Physics, The Pennsylvania State University, University Park, Pennsylvania 16802, USA}
	\affiliation{Department of Engineering Science and Mechanics, The Pennsylvania State University, University Park, Pennsylvania 16802, USA}
	
	
	\begin{abstract}
		Physical systems evolve from one state to another along paths of least energy barrier. Without \textit{a priori} knowledge of the energy landscape, multidimensional search methods aim to find such minimum energy pathways between the initial and final states of a kinetic process. However in many cases, the user has to repeatedly provide initial guess paths, thus ensuring that the reliability of the final result is heavily user-dependent. Recently, the idea of ``distortion symmetry groups" as a complete description of the symmetry of a path has been introduced. Through this, a new framework is enabled that provides a powerful means of classifying the infinite collection of possible pathways into a finite number of symmetry equivalent subsets, and then exploring each of these subsets systematically using rigorous group theoretical methods. The method is shown to lead to the discovery of new, previously hidden pathways for the case studies of bulk ferroelectric switching and domain wall motion in proper and improper ferroelectrics, as well as in multiferroic switching. This approach is applicable to a wide variety of physical phenomena ranging from structural, electronic and magnetic distortions, diffusion, and phase transitions in materials.
	\end{abstract}
	
	\maketitle
	
	\section{Introduction}

	\begin{figure*}[t]
		\centerline{\includegraphics[width=\linewidth]{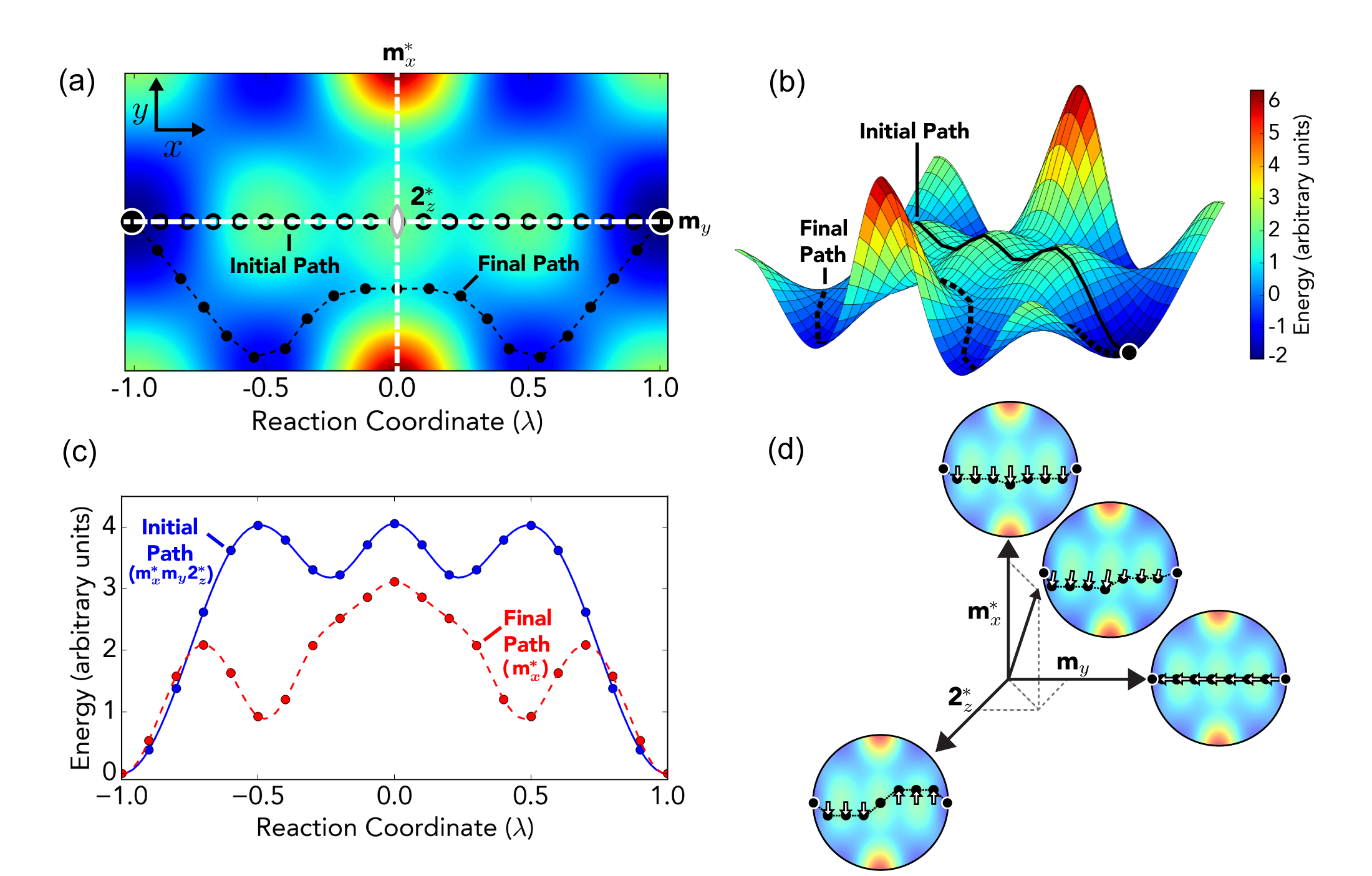}} 
		\caption{(a)(b) The potential energy landscape and images of atoms representing atomic motion on the surface of a material. The initial linear path has a path symmetry of $m_x^* m_y^* 2_z^*$. Perturbing the initial path with the symmetry-adapted perturbation associated with the $\Gamma_2$ irrep brings the path down to $m_x^*$ symmetry, and produces the lowest energy path once an NEB calculation is run. Both the initial linear path and the final relaxed path resulting from the NEB algorithm are shown. (c) The energy relative to the initial and final state as a function of reaction coordinate for both the high (initial) and low energy paths. (d) The symmetry adapted components projected out of an arbitrary perturbation to the initial path. The mode shown along the  axis is used to perturb the initial high-energy path. \label{fig1}}
	\end{figure*}

	Minimum energy paths (MEPs) \cite{Hratchian2005} are of utmost importance in physical sciences where they are used to study any arbitrary distortion in a material, such as changes in molecular confirmations and chemical reactions \cite{Pechukas1981,Bone1992,McIver1972}, structural or electronic phase transitions \cite{Zhuang2017,Wang2017,Lu2016}, diffusion \cite{Shang2011,Fang2014,McIver1972}, ferroelectric and magnetic switching \cite{Huan2014,Demaske2016,Bennett2013,Heron2014,Nowadnick2016}, surface reconstructions and the motion of interfaces such as dislocations \cite{Vegge2000,Shang2014} and domain walls \cite{Kumagai2013,Chandrasekaran2013}. All of these processes require knowledge of the underlying atomic motion provided by the MEPs. Finding MEPs can however be difficult due to the lack of \textit{a priori} knowledge of the potential energy landscape, given the prohibitive computational cost in determining it for processes involving a large number of atoms. The goal is then, given a starting and ending point in the landscape, to find the path connecting them that provides the lowest energy barrier, such that the forces perpendicular to the path are at a minimum; this defines an MEP. 
	
	Among chain-of-states methods for determining MEPs \cite{Tao2012}, a commonly used approach is with nudged elastic band (NEB) calculations \cite{JONSSON1998}. However, the application of this method raises a major difficulty; the relaxed path depends necessarily on the initial conditions, and repeated multiple starting points are required to explore additional pathways that may exist along the potential energy surface. Traditionally, this initialization is performed stochastically, or by altering one or several images to construct new paths via physical intuition or geometry optimization \cite{Olsen2004,Zimmerman2013,Schlegel2011,Jafari2017}. While this is effective in many situations given sufficient computational attempts, it does not treat the path as a singular unit, and as will be shown, can lead to the omission of lower energy paths. Further, this approach cannot ensure that the path found is indeed the one of lowest energy.
	
	These limitations of the NEB algorithm have recently been formally described by Vanleeuwen and Gopalan, using the language of symmetry \cite{VanLeeuwen2015}. They show that the symmetry of a distortion pathway in a material is entirely described by one of the 17,803 double antisymmetry space groups \cite{Vanleeuwen2014}, and that this symmetry is only conserved or raised by the NEB algorithm, but never lowered. It is this fundamental fact that restricts access to MEPs with lower path symmetry without a symmetry-breaking perturbation of the initial path. In contrast to stochastic methods of breaking path symmetry in which only one or a few intermediate images are perturbed, as is currently practiced, group theory offers a systematic way to perturb an entire path (including all the images) using modes obtained from irreducible representations (irreps). By perturbing with these modes, the symmetry of the path can then be systematically lowered. This symmetry-based method thus provides a more comprehensive and efficient stepwise procedure for classifying all of the infinite paths into a finite number of symmetry equivalent subsets, and then exploring each of these using group theoretical tools.
	
	To motivate the need for symmetry-adapted perturbations, consider the potential energy landscape shown in Figures \ref{fig1}(a) and \ref{fig1}(b). (For the sake of illustration, the landscape here is given \textit{a priori}, but the conclusions of the analysis are general.) If this system is taken to represent the motion of an atom on the surface of a material, a symmetry analysis of the path can be performed using only the information provided by the potential energy landscape. In this example, the reaction coordinate, $\lambda$, is naturally taken as the horizontal coordinate which varies between the two states, renormalized from $-1$ (the initial state) to $+1$ (the final state). To describe path symmetry, a new antisymmetry operation called distortion reversal symmetry (1*) was introduced by VanLeeuwen and Gopalan \cite{VanLeeuwen2015} who’s action on $\lambda$ is as follows: 1*($\lambda$) = $-\lambda$. Taking the initial path as the linear interpolation of images between the two end points, the resulting distortion symmetry is $m_x^* m_y 2_z^*$ (see Supplementary Note 1 for its determination). The notation $m_y$ for example represents a mirror perpendicular to the $y$-axis, and $2_z$ represents a 2-fold rotation axis parallel to the $z$-axis as shown in Figure \ref{fig1}. The three non-trivial operations in this group are $m_x^*$, $m_y$, and $2_z^*$, which leave the initial path shown in Figure \ref{fig1} invariant when applied to all of the images that compose it. Operations marked with an asterisk are referred to as ``starred" operations, and can be physically interpreted as conventional spatial operations that also require reversing the direction of the path in order to leave it invariant. It should be noted that the methods described in this paper can always be applied, irrespective of whether starred symmetry operations are present or not.  Furthermore, even if the MEP lacks any significant symmetry, it is sometimes possible to construct a highly symmetrized initial starting path that allows us to discover the MEP using the proposed group theoretical methods.
	
	Running an NEB calculation on the initial linear path with distortion group $m_x^* m_y 2_z^*$ produces the energy profile shown in blue in Figure \ref{fig1}(c). However, in this example, an alternate path with a lower energy barrier exists that is initially inaccessible due to the balanced forces arising from path symmetry conservation in the NEB algorithm. To obtain this lower energy path, the initial linear path needs to be perturbed. Group theory shows that any arbitrary perturbation of this initial path can be decomposed into a linear combination of \textit{exactly} three non-trivial symmetry-adapted perturbative components (Figure \ref{fig1}(d)) that transform as the three irreps, $\Gamma_2$, $\Gamma_3$, and $\Gamma_4$ shown in Supplementary Table 1. These three irreps have kernel symmetry groups of $m_x^*$, $m_y$, and $2_z^*$ respectively. Modifying the initial path with symmetry lowering perturbations such that the new path transforms as one of these kernel symmetry groups, results in three new initial paths that conserve only these symmetries. The new initial path with $m_y$ symmetry simply returns to the original linearly interpolated path once the NEB algorithm is employed. It is only the $m_x^*$ and $2_z^*$ perturbations that produce new paths after NEB optimization, with only the $m_x^*$ path providing a path with a lower overall energy barrier. Thus, the MEP in this example has a path symmetry group of $m_x^*$.
	
	Next, we illustrate the power of the above method using three examples of ferroelectric polarization switching, two involving bulk switching, and one involving a domain wall motion.  Understanding the minimum energy pathway for polarization reversal is of practical interest to a broad range of technologies such as optical communications, ultrasound imaging and sensing, precision actuation, infrared imaging and nonvolatile memory.  While bulk ferroelectric switching provides a simplistic picture of the polarization reversal process that often involves defects and grain boundaries, it is computationally less intensive, and still provides important insights into the atomic behavior in an ideal lattice.  Further, the symmetry method described below can also be used for larger molecular dynamics simulations of domain wall motion using symmetrized supercells \cite{Liu2016}.

	\section{Methods}
	Nudged elastic band (NEB) calculations \cite{JONSSON1998} were used in this study. The Vienna Ab initio Simulation Package \cite{Kresse1993,Kresse1996a,Kresse1996,Kresse1999} was used for all structural optimization and NEB calculations. The ability to generate symmetry adapted perturbations was implemented into a standalone piece of Python code and the open-source Quantum-ESPRESSO \cite{Giannozzi2009} software package. This is discussed in more detail in the implementation section of the supplementary material. VESTA \cite{Momma2008} was used for the visualization of all of the structures.
	
	To generate the symmetry-adapted perturbations, a random perturbation to the path was first generated. Then, projection operators \cite{Bradley2010} were constructed using the matrix representations of the elements in the path’s distortion group, as well as the matrices of its physically irreducible representations. The former was obtained using the SPGLIB \cite{A.2009} library, and the latter with the listing by Stokes and coworkers \cite{Stokes2013}. The operators were then applied to a vector representing the arbitrary path perturbation to project out its symmetry-adapted components. The projection operators are defined as follows:
	
	\begin{equation}
	\hat{P}_{kk}^{\Gamma_n} = \dfrac{l_n}{h} \sum_{R}\left[D_{kk}^{\Gamma_n}\left(R\right)\right]^{*}\hspace{-0.1cm}\hat{R}
	\end{equation}
	
	Where $l_n$ is the dimension of the irrep $\Gamma_n$, $h$ is the order of the distortion group, $\hat{R}$ is a symmetry operation from the distortion group, and $D_{kk}^{\Gamma_n}\left(R\right)$ is the $k$th diagonal matrix entry of the matrix representation of element $R$ in the physically irreducible representation $\Gamma_n$. For irreps of dimension greater than one, multiple perturbative modes are obtained for each $k=1,...,l_n$. These form a representation space, where their linear combination is taken to produce the final perturbation.
	
	\subsection{Ca$_3$Ti$_2$O$_7$ calculations}
	A Z = 4 orthorhombic cell was used for the initial and final states. Structural optimization of these was completed using the PBEsol functional \cite{Perdew2008}, with a 6x6x2 $k$-point mesh, and a \SI{600}{\electronvolt} plane-wave cutoff. The projector augmented wave method was used to represent the ionic cores. There were 10 electrons for Ca ($3s^23p^64s^2$), 12 electrons for Ti ($3s^23p^64s^23d^2$), and 6 electrons for O ($2s^22p^4$) treated explicitly. NEB calculations were run using the G-SSNEB algorithm \cite{Sheppard2012}, until forces were below \SI{0.02}{\electronvolt\per\angstrom}. The path perturbations were normalized such that the maximum displacement of any one atom was set to \SI{0.1}{\angstrom}. 
	
	\subsection{BiFeO$_3$ calculations}
	A Z = 8 pseudocubic cell was used for the initial and final states (Supplementary Figure 2). Structural optimization of these was completed using the generalized gradient approximation (GGA) to density-functional theory with a 4x4x4 $k$-point mesh, and a \SI{500}{\electronvolt} plane-wave cutoff. The projector augmented wave method was used to represent the ionic cores. There were 15 electrons for Bi ($5d^{10}6s^26p^3$), 14 electrons for Fe ($3p^63d^64s^2$), and 6 electrons for O ($2s^22p^4$) treated explicitly. NEB calculations were run using the G-SSNEB algorithm \cite{Sheppard2012}, until forces were below \SI{0.02}{\electronvolt\per\angstrom}. The path perturbations were normalized such that the maximum displacement of any one atom was set to \SI{0.1}{\angstrom}. Magnetization data was obtained using the LSDA+U method \cite{Anisimov1997} to better treat the 3d electrons in Fe. Values of U = 4.0 and J = 0.5 were used \cite{Picozzi2009}.

	\subsection{PbTiO$_3$ calculations}
	Structural optimization of the structures of the initial and final states (Figure 4a) was completed using the PBEsol functional \cite{Perdew2008}, with a 1x6x6 $k$-point mesh, and a \SI{400}{\electronvolt} plane-wave cutoff. The projector augmented wave method was used to represent the ionic cores. There were 4 electrons for Pb ($6s^26p^2$), 4 electrons for Ti ($3d^34s^1$), and 6 electrons for O ($2s^22p^4$) treated explicitly. NEB calculations were run using the regular NEB algorithm, until forces were below \SI{0.01}{\electronvolt\per\angstrom}. The path perturbations were normalized such that the maximum displacement of any one atom was set to \SI{0.1}{\angstrom}.

	\section{Results and Discussion}
	\subsection{Application to domain switching in improper ferroelectric Ca$_3$Ti$_2$O$_7$}
	
	The first example involves the bulk switching of a recently discovered ferroelectric crystal Ca$_3$Ti$_2$O$_7$. This example illustrates how the commonly employed method of altering the initial path through physical considerations of the kinetic process may fail to discover hidden MEPs, especially when the images altered are shared by the new MEP. The example will also illustrate how to systematically explore distortions of the initial path constructed with both $k=(0,0,0)$ and $k\neq(0,0,0)$ irreps using group theory, where $k$ indicates a reciprocal lattice vector in the first Brillouin zone.

\begin{figure*}
		\centerline{\includegraphics[width=\linewidth]{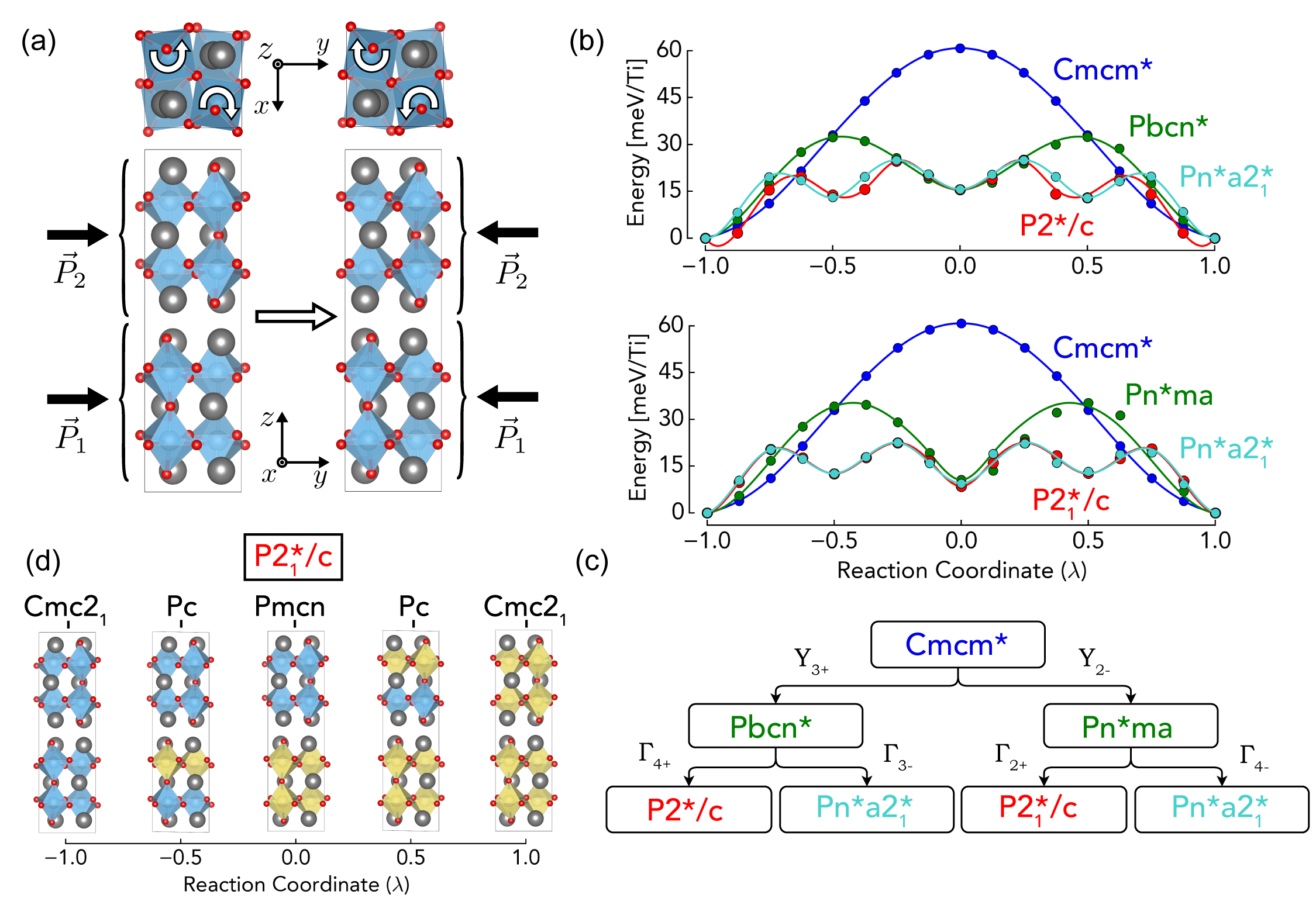}} 
		\caption{(a) The initial and final states of bulk polarization switching in the Ca$_3$Ti$_2$O$_7$ Ruddlesden-Popper ($n = 2$) structure. Arrows indicating the direction of the octahedral rotations in the $z$-direction are shown, as well as the polarization vectors for each of the perovskite slabs. Ca, Ti, and O are indicated by the grey, blue, and red atoms respectively. (b) The energy relative to the initial and final state as a function of reaction coordinate for both the two-step and four-step paths from the NEB method. The upper plot corresponds to perturbations illustrated by the left half of the tree in panel c, and the bottom with the right half. (c) Tree of distortion symmetry groups of resulting paths after perturbation. (d) A visual illustration of one of the four-step paths ($P2_1^*/c$) showing which sections of the perovskite slabs have transitioned from the initial to the final state. The spatial symmetry of each structure is shown above it. The two and four-step paths share the same transition image, with different lower symmetry images for the rest of the path. For example, in the $P2_1^*/c$ path, the intermediate images all have a spatial symmetry of $Pc$, whereas in the $Pn^*ma$ path, these have a symmetry of $Pma2_1$. \label{fig2}}
	\end{figure*}

	Inversion symmetry breaking in Ca$_3$Ti$_2$O$_7$ arises from a trilinear coupling between a polar mode and two other oxygen octahedral rotation modes \cite{Benedek2011}. Nowadnick and Fennie \cite{Nowadnick2016} employed various group theory methods in conjunction with physical intuition supported by density-functional theory (DFT) calculations to identify the switching pathway between a polarization ``up" and a polarization ``down" state, each with a structural symmetry of $Cmc2_1$ (see Figure \ref{fig2}(a)). Two low energy paths were identified that pass through intermediate structures of $Pnma$ and $Pbcn$ symmetry. Symmetry-adapted perturbations can be applied to this problem with significant results. First, a linearly interpolated path is constructed with a distortion symmetry of $Cmcm^*$ that passes through the high-symmetry paraelectric structure $Cmcm$. Using the NEB algorithm, the high-energy profile shown in blue in Figure \ref{fig2}(b) is revealed. From here, path perturbations are constructed using the non-trivial irreps of the group $Cmcm^*$ at the $\Gamma$ point $k=(0,0,0)$, and at the high symmetry $Y$ point $k=(1,0,0)$ (Supplementary Table 2). This latter point was chosen for this particular example, as the isotropy subgroups of the irreps appear to match the distortion groups of the previously reported low energy two-step paths (Figure \ref{fig2}(b)). In general, this knowledge would not be present, and perturbations would be generated and applied using irreps at all high symmetry $k$-points of the distortion group. It is important to note that perturbations constructed with these require a sufficiently large supercell in order to accommodate a loss of translational symmetry. After applying the perturbations, NEB calculations are run with two paths of lower energy being produced from the initial paths associated with irreps $Y_{4+}$ and $Y_{2-}$. These have $Pn^*ma$ and $Pbcn^*$ (Figure \ref{fig2}(c)) symmetry respectively, and match the low energy two-step paths reported by Nowadnick and Fennie \cite{Nowadnick2016}. The curves fitted to much of the data in this study, such as those in Figure \ref{fig2}(b), have been made symmetric about $\lambda = 0$ when the paths contains ``starred" symmetry. The asymmetries that are present in the data itself are as a result of numerical errors, which can commonly be seen in many reported MEPs (see Supplementary Table 1 in Ref. \cite{VanLeeuwen2015}).

	With previously reported paths having been recreated, further perturbations can be generated using the irreps of their distortion groups ($Pbcn^*$ and $Pn^*ma$). These are then applied to each respective path, and new NEB calculations are run, resulting in four new four-step paths of lower energy that were not previously reported. All of these have very similar barriers, with the $P2_1^*/c$ and $Pn^*a2_1^*$ paths from perturbation of the $Pn^*ma$ path having slightly lower energy barriers (\SI{22}{\milli\electronvolt}/Ti) as illustrated in Figure \ref{fig2}(b). Snapshots of the static structure at different points along the path for the $P2_1^*/c$ path are also displayed in Figure \ref{fig2}(d). The difference between the two-step and four-step paths can be seen in how the polarization in each layer ($\vec{P}_1$ and $\vec{P}_2$ in Figure \ref{fig2}(a)) changes. In the two-step paths, all four oxygen octahedra in each layer are rotated about the z-axis simultaneously, whereas in the four-step paths, each separate layer of octahedra rotate independently instead (Figure \ref{fig2}(d)). Ultimately, this results in low energy intermediate structures that were previously at the highest energy point of the two-step paths. This practical example illustrates the importance of the holistic treatment of NEB paths and their perturbations. By only considering different paths defined by alteration of the original transition image at $\lambda = 0$, additional lower energy paths may be missed such as in this example.
	
    \begin{figure*}	
	\centerline{\includegraphics[width=\linewidth]{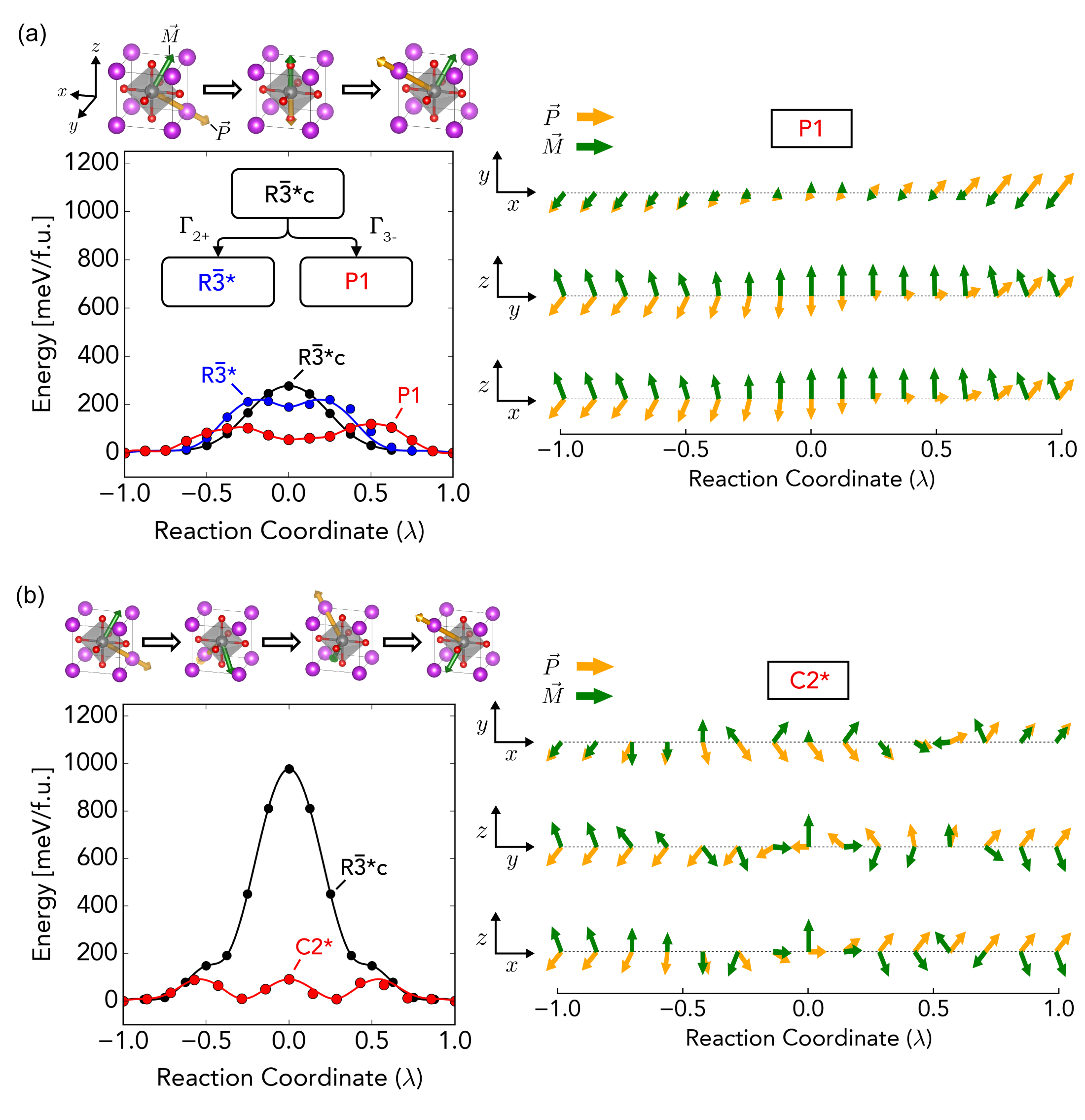}} 
	\caption{ (a) Data for the switching pathway without reversal of the octahedral rotations where \textit{\textbf{P}} is switched, and \textit{\textbf{M}} is not. Shown on the left is the energy relative to the initial and final states from NEB calculations as a function of reaction coordinate for all of the paths obtained from perturbation of the initial  path. Initial, middle, and final snapshots of \textit{\textbf{P}} and \textit{\textbf{M}} overlaid on a high-symmetry pseudocubic unit cell of BiFeO$_3$ is also displayed for the lowest energy $P1$  path. The purple, grey and red atoms represent Bi, Fe, and O respectively. Shown on the right are projections of \textit{\textbf{P}} and \textit{\textbf{M}} on various planes in the pseudocubic coordinate system for each image along the lowest energy path. (b) Data for the switching pathway with reversal of the octahedral rotations where both \textit{\textbf{P}} and \textit{\textbf{M}} are switched. Shown on the left is the energy relative to the initial and final states from NEB calculations as a function of reaction coordinate for all of the paths obtained from perturbation of the initial  path. Snapshots of \textit{\textbf{P}} and \textit{\textbf{M}} overlaid on a high-symmetry pseudocubic unit cell of BiFeO$_3$ are also displayed for the four lowest energy images in the  path. Shown on the right are projections of \textit{\textbf{P}} and \textit{\textbf{M}} on various planes in the pseudocubic coordinate system for each image along the lowest energy path. \label{fig3}}
    \end{figure*}
		
	\subsection{Application to multiferroic switching in BiFeO$_3$}
	The second example involves a well-studied multiferroic, BiFeO$_3$, and illustrates how distortion symmetry can help access and manipulate functional properties of materials such as coupling of polarization and magnetization. BiFeO$_3$ is of interest for its well-known room temperature magnetoelectric coupling \cite{Yang2015}.  It exhibits a large ferroelectric polarization, \textit{\textbf{P}} (\SI{\sim100}{\micro\coulomb\per\centi\meter\squared}) \cite{Picozzi2009} and a weak ferromagnetic moment, \textit{\textbf{M}} ($\sim0.1-1.0$ $\mu_B/Fe$) arising from the canting of the antiferromagnetically aligned spins through an antisymmetric Dzyaloshinskii-Moriya (DM) interaction \cite{Ederer2005}. The interest is in a \textit{deterministic} control of the magnetization \textit{\textbf{M}} using an electric field to control the polarization, \textit{\textbf{P}}, as recently reported in literature \cite{Heron2014}. Should we expect a deterministic control of \textit{\textbf{M}} with controlling \textit{\textbf{P}}? At least in the bulk switching case, we discover a new energetically competing pathway that is close in energy to the previously reported path, and thus prevents such determinism.  We also show that this pathway proceeds through a configuration that breaks the mutual orthogonality of \textit{\textbf{M}} and \textit{\textbf{P}}, contrary to what is normally assumed.

	To study this interaction, we start by considering two possible pathways to switch \textit{\textbf{P}} - a deterministic path with octahedral rotations switching between the initial and final states, and a nondeterministic path without such switching of octahedral rotations. By doing so, we are able to specify whether \textit{\textbf{M}} is also switched in the process, since a reversal of the oxygen octahedra necessitates a switching of the DM vector, and hence \textit{\textbf{M}} as well \cite{Heron2014}. Data for the two different switching pathways are shown in Figures \ref{fig3}(a) and \ref{fig3}(b), respectively. For the path in panel a, only the polarization is reversed (nondeterministic), whereas for the path in panel b, the octahedral rotations and DM vector are both reversed (deterministic). Constructing both paths using a linear interpolation between the end states shown in Supplementary Figure 2 results in path symmetries of $R\bar{3}^*c$.  By performing NEB calculations, the black energy profiles shown in Figures \ref{fig3}(a) and \ref{fig3}(b) are produced. At this stage of exploration, additional NEB calculations can be performed after applying symmetry adapted perturbations constructed from the irreps of $R\bar{3}^*c$ shown in Supplementary Table 3. What is ultimately obtained is a set of two paths with much lower overall energy barriers than the original, resulting from perturbations that transform as the $\Gamma_{3-}$ irrep. It should be noted that in the case of the lowest energy path in panel b, the final path symmetry is $C2^*$ and not $P1$. While this is not the direct kernel of the irrep, it is one of its epikernels, and thus the NEB algorithm raising path symmetry to it is not surprising (see Supplementary Note 1 in Ref. \cite{VanLeeuwen2015}). For the deterministic path in b, the MEP has an energy barrier of \SI{90}{\milli\electronvolt}/f.u. and consists of three $71^\circ$ switching steps for \textit{\textbf{P}} as it transitions from $[\bar{1}\bar{1}\bar{1}]_{ps}$ to $[111]_{ps}$, while passing through $[1\bar{1}\bar{1}]_{ps}$ and $[1\bar{1}1]_{ps}$. For the nondeterministic path in a, the energy barrier is \SI{120}{\milli\electronvolt}/f.u., and the resulting path exhibits two-step coordinated sequential motion of half of the Bi atoms and oxygen octahedra, producing the notable polarization and magnetization vectors along the path shown in Figure \ref{fig3}. In the first step of the nondeterministic path, half of the Bi atoms move along $[111]_{ps}$, while half of the octahedra rotate, causing all of them to be in phase about $[001]_{ps}$. In Glazer notation, this corresponds to a transition from an $a^-a^-a^-$ to $a^-a^-a^+$ structure. In the second half of the path, the octahedra return to their initial state, as the other half of the Bi atoms move along $[111]_{ps}$. This sequential motion of the \textit{A} cations is of note, as recently this kind of behavior has been shown to lower bulk ferroelectric switching energies in corundum structured materials \cite{Ye2016}. One additional difference between both paths is the behavior of \textit{\textbf{M}}. In the deterministic path, \textit{\textbf{M}} and \textit{\textbf{P}} stay perpendicular to each other throughout the switching process. In the nondeterministic path however, \textit{\textbf{M}} transitions from a perpendicular to parallel configuration relative to \textit{\textbf{P}} to lower the overall energy barrier. This is of note, as it deviates from conventional assumptions for bulk BiFeO$_3$ about the perpendicular relationship between \textit{\textbf{M}} and \textit{\textbf{P}} that is exhibited in the deterministic path.
	
	With both MEPs being of similar overall energy barrier, two competing pathways are presented – one which results in deterministic switching of \textit{\textbf{M}} (\SI{90}{\milli\electronvolt}/f.u.), and one that does not (\SI{120}{\milli\electronvolt}/f.u.). Previous studies on strained BiFeO$_3$ have missed the latter path which is only \SI{30}{\milli\electronvolt}/f.u. higher, on the order of room temperature (\SI{25}{\milli\electronvolt}) \cite{Heron2014}. This study reveals a more nuanced picture that is more consistent with the experimental data (see supplementary data in Ref. \cite{Heron2014}) that shows switching proceeding through both pathways in the strained material, indicating direct competition between these paths. Knowledge of such competing pathways can now be used to inform decisions about materials design. If one pathway is desired more than the other, it is possible that perturbations such as domain walls and strain could be utilized to increase its favorability.

    \begin{figure*}[t]
    	\centerline{\includegraphics[width=\linewidth]{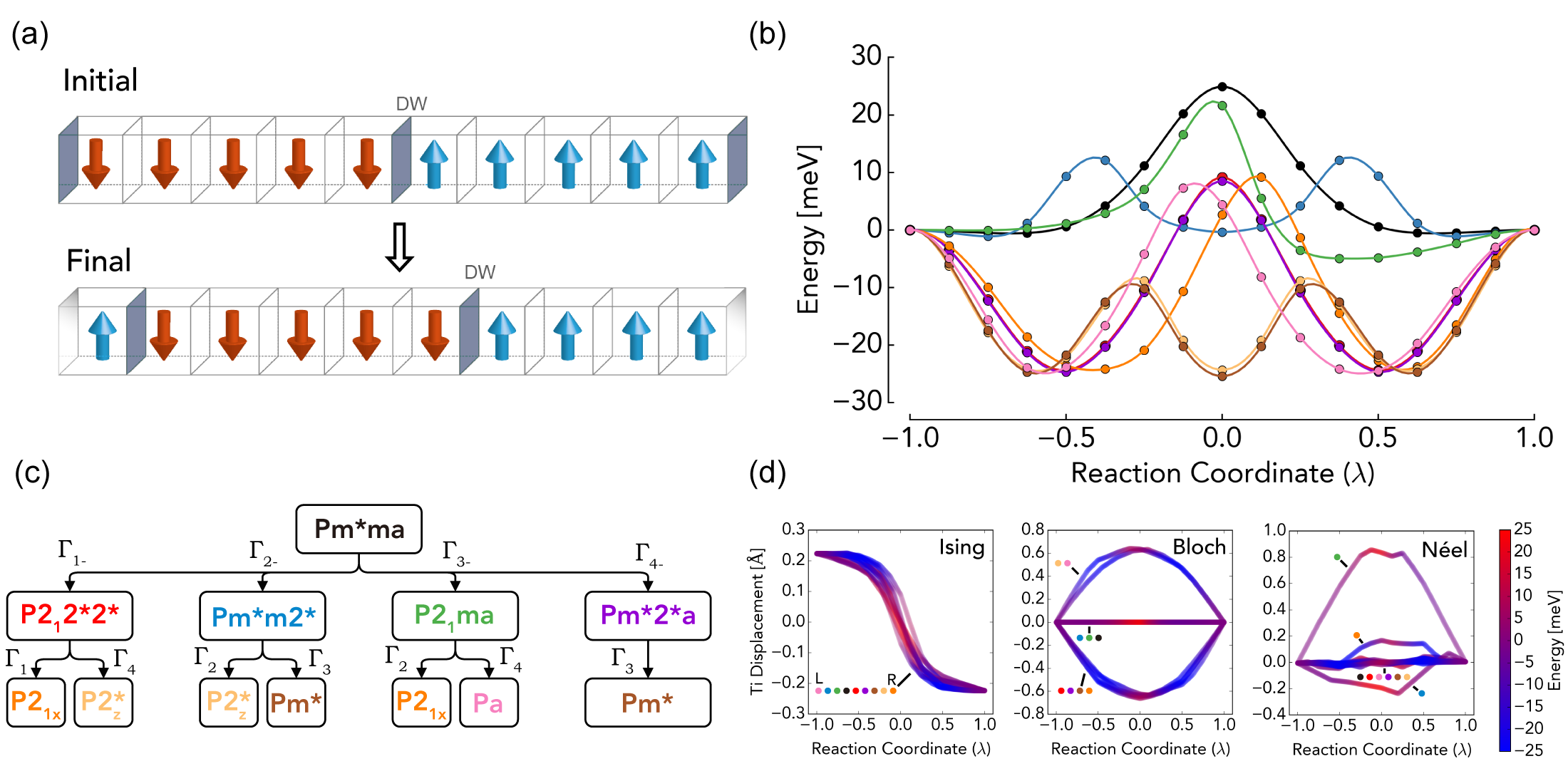}} 
    	\caption{ (a) The structure of the supercells of the initial and final states used to construct the initial path for $180^\circ$ domain-wall motion. Each box indicates a PbTiO$_3$ unit cell, with the red and blue arrows indicating polarization direction. (b) The energy relative to the initial and final states as a function of reaction coordinate for the final paths obtained from NEB calculations. The colors indicate the distortion group of the final path shown in the tree in panel c. (c) The tree of distortion symmetry groups resulting from path perturbations. (d) The Ising, Bloch, and N\'eel components of the Ti displacement for the first unit cell in the structures shown in panel a as a function of reaction coordinate. The energy of the complete structure is indicated by the color of the line, with the colored labels indicating which paths are present in the data. \label{fig4}}
    \end{figure*}

	\subsection{Application to domain wall motion in ferroelectric PbTiO$_3$}
	The third example for this method illustrates the possibility for exploration of the potential energy landscape by choosing an initial highly symmetrized path for the motion of a domain-wall in PbTiO$_3$ (Figure \ref{fig4}(a)). 
	
	$180^\circ$ ferroelectric domain-walls in PbTiO$_3$ have been studied extensively using a variety of first-principles techniques \cite{Lee2009,Behera2011,Wang2013,Wojde2014,Wang2014,Meyer2002,He2003}. Although the PbTiO$_3$ supercell with Pb-centered domain walls used to construct the initial and final states in this study (Figure \ref{fig4}(a)) was relaxed, the true ground state structure was not initially obtained due to the constraining symmetry of the static structure. While this may result in paths with images that are lower in energy relative to the end points, it will be shown that this allows for an exploration of some of the neighboring energy landscape, and in turn, some of the many ways in which the different components of the polarization can behave during domain-wall motion. 
	
	First, an initial NEB calculation was carried out using the linearly interpolated path between the initial and final states. This has a path symmetry of $Pm^*ma$, and the resulting energy profile is shown in Figure \ref{fig4}(b). By constructing new initial paths for the NEB algorithm with symmetry-adapted perturbations, eight different final relaxed paths are obtained, as shown in Figures \ref{fig4}(b) and \ref{fig4}(c). By plotting the Ising, Bloch, and N\'eel components of the Ti displacement of one of the unit cells in the supercell through which the domain wall travels (i.e., the first cell in the structures in Figure \ref{fig4}(a)) as a function of reaction coordinate ($\lambda$), we can examine differences between the paths. The initial path shows the simultaneous movement of both walls in the supercell via a reduction of the Ising component of the polarization. For the $\Gamma_{2-}$ path, the pathway consists of the same type of polarization behavior, but with a sequential movement of each wall. For the other paths obtained, Bloch and N\'eel components arise in the pathways, as shown in Figure \ref{fig4}(d). Although many of these paths share similar magnitudes of these, each path with a different symmetry exhibits different underlying atomic motion. The actual ground state structure and the MEP are close to the relaxed lowest energy structures and the path between them seen in Figure \ref{fig4}(b); see Supplementary Figure 6 for details.

	\section{Conclusion}
	In this work, we have illustrated a powerful new symmetry-based approach to discovering new MEPs. By utilizing group theoretic tools, paths can be categorized according to their symmetry, and symmetry-adapted perturbations can be generated to allow this symmetry to be broken in a systematic manner. Using this approach within the NEB method, subsets of paths can be defined, each of which is explored for new MEPs. By applying this technique to a variety of systems involving bulk ferroelectric switching and domain-wall motion, new insights were gained. Four-step bulk switching pathways were discovered for the improper ferroelectric Ca$_3$Ti$_2$O$_7$ with a large unit cell and a complex domain pattern. A competing new pathway was discovered for multiferroic switching in BiFeO$_3$ that allows for switching of polarization without switching canted magnetism, resulting in an indeterministic multiferroic switching process that competes with a deterministic path; this result informs studies where deterministic switching is sought \cite{Heron2014}. A rich range of domain wall motion paths in the energy landscape of PbTiO$_3$ were discovered starting from a highly symmetrized initial path; walls with mixed Ising, Bloch, and N\'eel character were observed during the switching process. 
	
	In order to complete calculations for the examples in the study, the distortion symmetry-based method was implemented to interface with standard NEB codes. This has resulted in an observed speedup of NEB calculations on paths containing starred symmetry of up to two times (see Supplementary Figures 7 and 8). Furthermore, as demonstrated in this work, even in cases when the final MEP may have low symmetry, one can often start with a highly symmetrized initial pathway that allows the user to exploit the power of this method. In the future, we envision the possibility for further extensions to the distortion symmetry framework, by incorporating other kinds of symmetry groups, such as those involving distortion translation \cite{Padmanabhan2017}. We foresee that the newly proposed symmetry-adapted framework could become an integral part of the discovery of new transition states in many problems of relevance to materials science.
	\\
	\section*{Acknowledgments}
	We gratefully acknowledge support from the NSF-MRSEC Center for Nanoscale Science at the Pennsylvania State University, Grant \#DMR-1420620, and the JSPS KAKENHI, Grant \#JP16H06793 and \#JP17K19172, and the Murata Science Foundation. J.M.M. and I.D. also acknowledge partial support from the Soltis faculty support award and the Ralph E. Powe junior faculty award from Oak Ridge Associated Universities. We also appreciate Prof. F. Oba for providing his computational resources.

\bibliography{ref}
	
\end{document}